**Aztec: A Platform to Render Biomedical Software Findable, Accessible, Interoperable, and Reusable**

Wei Wang[1,2]*, Brian Bleakley[2,3], Chelsea Ju[1,2], Vincent Kyi[2,3], Patrick Tan[1,2], Howard Choi[2,3], Xinxin Huang[1,2], Yichao Zhou[1,2], Justin Wood[1,2], Ding Wang[2,3], Alex Bui[2,4], Peipei Ping[2,3,5]*

[1]Department of Computer Science, Henry Samueli School of Engineering and Applied Science, UCLA

[2]NIH BD2K Center of Excellence for Biomedical Computing at UCLA

[3]Department of Physiology, David Geffen School of Medicine, UCLA

[4]Department of Radiology, David Geffen School of Medicine, UCLA

[5]Department of Medicine, David Geffen School of Medicine, UCLA

*Correspondence: weiwang@cs.ucla.edu

*Correspondence: ppingucla@gmail.com

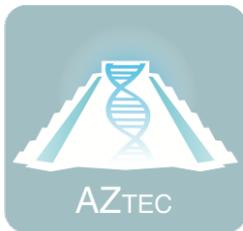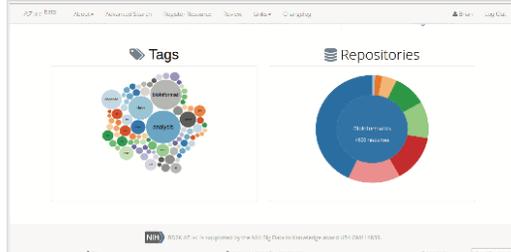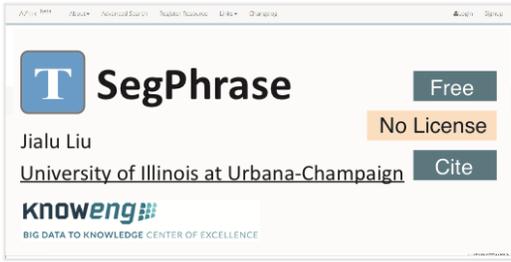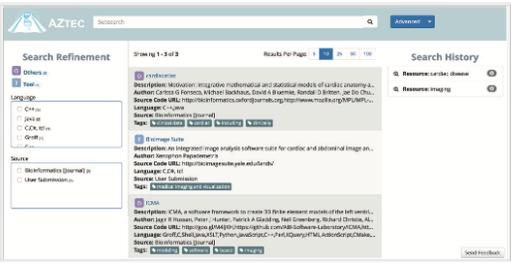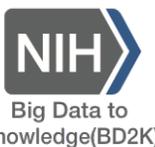

**In Brief:**

Aztec is a cloud-based online resource for investigators to find and access biomedical tools and resources encompassing 17 major domains, including omics, imaging, and machine learning.

**Highlights**

- Aztec is an online tool indexing resource
- An automated process identifies tools and extracts metadata from PubMed
- Aztec provides a unified hub for researchers to find tools
- Aztec extends the lifespan of tools by increasing their exposure

**Summary**

Precision medicine and health requires the characterization and phenotyping of biological systems and patient datasets using a variety of data formats. This scenario mandates the centralization of various tools and resources in a unified platform to render them Findable, Accessible, Interoperable, and Reusable (FAIR Principles). Leveraging these principles, Aztec provides the scientific community with a new platform that promotes a long-term, sustainable ecosystem of biomedical research software. Aztec is available at https://aztec.bio and its source code is hosted at https://github.com/BD2K-Aztec.

**Introduction**

Advances in bioinformatics, especially in the field of genomics, have greatly accelerated as development of powerful computational systems and new distributed platforms enabled rapid processing of massive amounts of datasets. Accordingly, this has led to an explosion of public software tools, databases, and other resources for biological discovery (Henry et al., 2014).

Modern biomedical research requires the comprehension and integration of multiple types of data and tools; specifically, understanding biomedical phenotypes requires analysis of molecular data (genomics, proteomics), imaging data (sonography, CT), and textual data (case reports, electronic health records) (Payne, 2012). Researchers require software tools in a common platform that can analyze and integrate data from each of these domains. However, many of the existing resource and tool repositories are narrowly focused, fragmented, or scattered and do not support the multidimensional data analysis required for advancing precision medicine (Cannata et al., 2005). There is a lack of unified platform collection software applications across various disciplines, leaving many users unable to locate the tools critical to their research.

To manage this influx of digital research objects (tools and databases), the FORCE11 community put forth a set of guiding principles to make data Findable, Accessible, Interoperable, and Reusable (FAIR) (Wilkinson et al., 2016). While these issues have been addressed in the data elements domain, they are equally important to but thus far neglected by the biomedical informatics software domain.

Here we introduce Aztec (**A** to **Z tec**hnology; https://aztec.bio), an open-source online software discovery platform that empowers biomedical researchers to find the software tools they need. Aztec currently hosts over 10,000 resources spanning 17 domains, including imaging, gene ontology, text-mining, data visualization, and various omics analyses (i.e., genomics, proteomics, and metabolomics). Aztec serves as a common portal for biomedical software across all of these domains, enabling modern multi-dimensional characterization of biomedical processes.

**Applying FAIR Principles to Biomedical Software**

Our goal is to build a platform for enabling and ensuring the compliance of software and tools with FAIR principles.

First, we address the issue of **Findability**. Datasets are rendered Findable by rich, accompanying metadata describing how datasets were generated and how they can be analyzed. There are specific functions that each software tool can fulfill, and each tool has its own restrictions and requirements regarding its application. For a software tool to be considered Findable, researchers must be able to query and filter based on tool function, biological domain, prerequisites, input and output formats, and a host of other vital metadata fields. Aztec will enable the querying of all of these features, thus eliminating obstacles for finding specific tools faced by researchers today. While many robust, metadata-enriched databases exist to index available biological datasets (Perez-Riverol et al., 2016), few comparable systems exist to index and describe tools (Ison et al., 2016).

Second, Aztec expands tool **Accessibility**. Though only the publisher of a tool can guarantee that it is accessible, Aztec eliminates many barriers of universal tool accessibility. Notably, Aztec's RESTful Application Program Interface (API) is accessible without a key and provides all users with complete tool metadata, including licensing, availability, and resolvable tool URL.

Third, Aztec addresses **Interoperability** by adopting existing vocabularies to define tool metadata whenever possible. This ensures that Aztec tool metadata is interoperable with a wide array of other APIs, online platforms, and analysis tools. Using established vocabularies from GitHub ([https://github.com](https://github.com)), EDAM ([http://edamontology.org](http://edamontology.org)), CrossRef ([http://www.crossref.org](http://www.crossref.org)), and BioPortal ([http://bioportal.bioontology.org](http://bioportal.bioontology.org)), Aztec describes each tool with rich metadata that can be queried and matched with metadata on other platforms (Table 1). Aztec returns all metadata queries using the well-established, industry-wide standard JSON Schema.

Finally, Aztec is committed to ensuring the **Reusability** of both the Aztec platform and the tools indexed within it. Aztec is a well-documented open source platform ([https://github.com/BD2K-Aztec](https://github.com/BD2K-Aztec)) released under the highly-permissible MIT License. Aztec's metadata offers detailed descriptions of the technical, legal, and commercial operating requirements for all indexed tools whenever available.

Leveraging these principles, Aztec provides the scientific community with a new platform that promotes a long-term, sustainable ecosystem of biomedical research software.

**Data Sources**

Aztec operates on an automated pipeline that collects, clusters, and indexes tools from four different sources:

(1) Publications. The Aztec database is populated predominantly by tools that are described in the biomedical literature indexed by PubMed. All new publications from journals of interest—those which regularly publish new biomedical software tools—are programmatically examined and classified as either positive or negative tool publications using a neural network classifier. Tool metadata is then extracted from the publication using the PubMed API, article abstract, article full text (where available), and the GitHub API (where resolvable GitHub URLs are found in the abstract or full text). Existing general-purpose biomedical software indexes rely on manual entry of new software tools (Henry et al., 2014; Ison et al., 2016). By automatically extracting metadata from PubMed, Aztec will feature new tools as soon as they are published online.

When retrieving tools from PubMed, Aztec can incorporate tools published in both open access journals, where the full text is public, and conventional journals, where only the abstract is freely available. Using the PubMed API, Aztec can automatically retrieve information on the tool's authors, their institutions, and Medical Subject Heading (MeSH, https://meshb.nlm.nih.gov) terms describing the tool's function (see Table 1). Additionally, Aztec can retrieve either the publication abstract or the full article in PDF format. Using GROBID, a machine learning library for parsing scientific literature, Aztec can extract the full text, citations, and acknowledgements from PDF publications (Lopez, 2009).

The metadata extracted from PubMed literature can be enriched significantly when the tool's source code is hosted in a public repository such as GitHub. Repository URLs are automatically identified by Aztec in article text and queried using the GitHub API. The API returns metadata such as tool name, programming language, and license (see Table 1). Additionally, GitHub's API returns important usage metrics such as forks and commits, which Aztec uses to rank search results. Where repository URLs are not available, tool names are extracted from publication titles using refined heuristics.

(2) <u>Other tool indexes</u>. Aztec also functions as a tool repository aggregator. Using both RESTful APIs and HTML parsers, Aztec extracts tool metadata from other bioinformatics tool indexes, including BioJS (https://biojs.net), Bioconductor (https://www.bioconductor.org), Cytoscape (http://www.cytoscape.org), and Sourceforge (https://sourceforge.net). In developing the Aztec metadata schema, existing index metadata schemas were adhered to as closely as possible to facilitate resource index interoperability and reusability.

(3) <u>User submissions</u>. Using a wiki-style collaborative editing system, users can make changes directly to the tool metadata in Aztec. This enables authors to complete empty fields that were missed by Aztec's metadata extraction system and to correct any errors. Users can also submit entirely new tools to Aztec's database, creating new entries for tools that were not identified in PubMed.

(4) <u>Funding sources</u>. For NIH IC-funded research, grant numbers are matched to IC names, enabling users to find all software funded by a particular IC. For example, in the case of the NIH BD2K initiative, grant numbers are used to map tools to specific NIH Centers of Excellence in Biomedical Big Data Computing (See Table 2).

**Platform Architecture**

Aztec is cloud-based and is hosted on an Amazon EC2 server, with the flexibility to scale up to meet user demand. Aztec is implemented with the Node.js framework using an enterprise-established Model-View-Controller design pattern. Users can interact with Aztec either through a web-based graphical interface or directly through an API.

Our platform employs a user management system with email authentication and SSL encryption to ensure secure resource registration and updates. Aztec offers an advanced yet user-friendly query system, allowing a cascade of nested sub-searches as well as comprehensive filtering and sorting of results. To preserve data consistency and integrity across diverse and voluminous inputs, Aztec employs a relational database implemented in MySQL. Powered by Aztec-IR, Aztec supports efficient keyword search and semantic query guided by established ontologies. On the web application's front-end, Aztec is implemented with the Bootstrap framework. It also incorporates D3 (https://d3js.org), a JavaScript graphing library that integrates HTML, SVG, and CSS to create interactive graphical visualizations.

**Aztec-IR**

Aztec is powered by Aztec-IR, our custom rich semantic Information Retrieval system, which outperforms the industry standard Apache Solr (http://lucene.apache.org/solr) enterprise search system when user queries can be mapped to a particular domain corpus. Aztec-IR first summarizes the information in raw text and then expands the query in a reasoned way within a particular ontology and knowledge hierarchy.

Aztec provides multiple search functions, including search by keywords, search by specific fields, and searching within results. To return accurate search results and provide a better user experience, Aztec also suggests search terms based on user queries in order to control the user input vocabulary. Our custom information retrieval system, Aztec-IR, is able to perform domain categorization tasks, generate reliable knowledge bases for query, and recommend similar tools (Figure S1). It consists of three main components that perform the following tasks:

(1) Building a thesaurus. The thesaurus is built from a corpus of documents (referred to as C1 in Figure 2) that are representative of the language used in the domain of interest; in Aztec, this corpus consists of the biomedical literature retrieved from PubMed. Synonyms are generated from semantic similarity analysis based on the corpus C1. Hypernyms and hyponyms are retrieved from the related biological ontologies (MeSH; EDAM, http://edamontology.org; NCIT, https://ncit.nci.nih.gov) to enrich the thesaurus.

(2) Converting documents into vectors. Aztec-IR processes the tool descriptions and metadata documents that will be retrieved in response to user queries. Each such document is transformed into a vector of dimensionality equal to the thesaurus length. User queries are transformed and expanded via the same process into a vector of the same dimension. Search results are generated by ranking the relevance of document vectors to query vector. Within groups of results with close similarity to the query vector, results are ranked by usage metrics retrieved from the GitHub API.

(3) Topic classification. We have also leveraged the accurate feature space created by Aztec-IR to represent documents for topical categorization. Categorization of tools within specific biological domains improves tool organization and the quality of information retrieval, as users can constrain the search of tools to a specific domain in order to find the desired ones more quickly. Each tool has a description with short unstructured text (referred to as C2 in Figure 2), which is leveraged to build document vectors using Doc2Vec converter (https://deeplearning4j.org/doc2vec; see Figure 2).

Our topic-classification model achieves 87.5% accuracy, outperforming other approaches such as TF-IDF and Labeled-LDA with 83.9% and 79.5% accuracy, respectively (Ramos, 2003) (Blei et al., 2003). Building a semantic rich feature space and capturing semantic similarities is important for yielding relevant search results, especially when the metadata is brief and sparse. This approach could be expanded to other platforms where search language can be matched to the controlled ontology of a specialized domain.

(4) Retrieving document. User queries are expanded and transformed using the same techniques for the documents. Search results are generated by ranking the relevance of document vectors to query vector. Within groups of results with close similarity to the query vector, results are ranked by usage metrics retrieved from the GitHub API.

**Indexing tools and resources**

Aztec uses two independent systems to uniquely identify tools in the database. Tools are tracked internally using a Resource Identifier (RID) consisting of the prefix 'AZ' followed by a sequential integer. These RIDs define the URLs that link to particular tools and will be displayed on the tool page, but they are not intended to be referenced in publications or other databases outside of Aztec. A Digital

Object Identifier (DOI) will be used to generate citations in a format consistent with the expectations of the academic publishing community. Tools with pre-existing DOIs, registered by an organization such as Zenodo (https://zenodo.org), will have that information stored in the tool's metadata.

**Other Ontologies and Identifiers**

Aztec will use the EDAM ontology for bioinformatics operations, data types, formats, identifiers, and topics. This ontology is used by the ELIXIR Tool and Data Service Registry (https://bio.tools/), as well as BioXSD (http://bioxsd.org), EMBOSS (http://emboss.sourceforge.net), eSysbio (http://esysbio.org) and SEQwiki (http://seqwiki.org). It is the most mature and robust existing ontology suitable for a bioinformatics tool registry. Additionally, EDAM has an online change request form and a GitHub repository that can be utilized to extend the ontology as needed to adequately describe all tools.

The Open Researcher and Contributor ID (ORCID, https://orcid.org) is a rapidly growing system to uniquely identify authors of scientific publications and datasets. At present, ORCID has almost 3 million registrants. The widespread adoption of this system will alleviate attribution issues stemming from the large number of researchers with similar names, as well as those whose names change or have multiple representations. Though not required, we recommend that Aztec contributors register an ORCID.

To enhance usability, Aztec automatically highlights terms in the tool description that match Medical Subject Headings (MeSH) and hyperlinks them to their entries in the BioPortal ontology database.

**Conclusion**

Aztec is a unique resource that provides an essential service to the biomedical research community. Through direct mining of the PubMed corpus, Aztec indexes new tools as they are first published, making them Findable by researchers well in advance of the time needed for manual entry into a similar tool database.

Further Aztec development is planned for the future. Analysis of journals will expand beyond *Bioinformatics* to include those that publish new tools less frequently and with less consistent formatting. Additionally, the metadata extraction methods will be continuously iterated and refined using both heuristic and machine learning methods.

As these new features are implemented, Aztec will become integral to propelling biomedical innovation in the digital era.

**Author Contributions**

BB, PP, and WW wrote the manuscript. BB, VK, CJ, PT, HC, JW, XH, and YZ developed the Aztec software, algorithms, and methods. DW contributed to graphic rendering of the platform. AB, BB, PP, and WW designed the Aztec platform and directed its execution.

**Acknowledgements**

This work was supported in part by NIH Awards U54GM114833 to WW and PP, as well as R35HL135772 to PP; and the UCLA Laubisch endowment, to PP.

**References**

Blei, D.M., Ng, A.Y., and Jordan, M.I. (2003). Latent dirichlet allocation. Journal of machine Learning research *3*, 993-1022.

Cannata, N., Merelli, E., and Altman, R.B. (2005). Time to organize the bioinformatics resourceome. PLoS Comput Biol *1*, e76.

Henry, V.J., Bandrowski, A.E., Pepin, A.S., Gonzalez, B.J., and Desfeux, A. (2014). OMICtools: an informative directory for multi-omic data analysis. Database (Oxford) *2014*.

Ison, J., Rapacki, K., Menager, H., Kalas, M., Rydza, E., Chmura, P., Anthon, C., Beard, N., Berka, K., Bolser, D.*, et al.* (2016). Tools and data services registry: a community effort to document bioinformatics resources. Nucleic Acids Res *44*, D38-47.

Lopez, P. (2009). GROBID: Combining Automatic Bibliographic Data Recognition and Term Extraction for Scholarship Publications. Proceedings of the 13th European Conference on Digital Library (ECDL), Corfu, Greece.

Payne, P.R. (2012). Chapter 1: Biomedical knowledge integration. PLoS Comput Biol *8*, e1002826.
Perez-Riverol, Y., Bai, M., Leprevost, F., Squizzato, S., Mi Park, Y., Haug, O.K., Carroll, A.J., Spalding, D., Paschall, J., Wang, M.*, et al.* (2016). Omics Discovery Index - Discovering and Linking Public Omics Datasets. bioRxiv.

Ramos, J. (2003). Using Tf-idf to Determine Word Relevance in Document Queries.

Wilkinson, M.D., Dumontier, M., Aalbersberg, I.J., Appleton, G., Axton, M., Baak, A., Blomberg, N., Boiten, J.W., da Silva Santos, L.B., Bourne, P.E.*, et al.* (2016). The FAIR Guiding Principles for scientific data management and stewardship. Sci Data *3*, 160018.

**Tables**

**Table 1: Aztec Metadata and Vocabulary Control**

| Field | Required | Ontology/Vocabulary |
|---|---|---|
| Name | Yes | |
| Accession number | Yes | Aztec Local Resource Identifier (LRI) |
| DOI | No | Digital Object Identifier (DOI) System |
| Description | Yes | |
| Logo | No | |
| Image(s) | No | |
| Link(s) | Yes | W3 URL Specification |
| Type | Yes | Locally defined |
| Function(s) | Yes | EDAM |
| Source repository(s) | No | W3 URL Specification |
| Language(s) | No | GitHub |
| Author(s) | Yes | ORCID |
| PI(s) | No | ORCID |
| Research Institution(s) | No | Wikidata |
| Primary publication | No | DOI |
| Other publication(s) | No | DOI |
| Funding source(s) | No | CrossRef |
| Award number(s) | No | |
| Biological domain(s) | Yes | EDAM |
| Release(s) | Yes | |
| Platform(s) | Yes | Locally defined |
| Input format(s) | No | EDAM |
| Output format(s) | No | EDAM |
| Upstream workflow tool(s) | No | Aztec LRI |
| Downstream workflow tool(s) | No | Aztec LRI |
| Reimplementation of | No | Aztec LRI |
| Reimplemented by | No | Aztec LRI |
| Submitter | No | |

**Table 1: Aztec Metadata and Vocabulary Control**

Rigorous vocabulary control has been implemented for many the critical metadata fields describing Aztec's tools. Some fields, such as "Platform(s)", we have standardized using our own vocabulary, focusing on simplicity and inclusivity. Many other fields have been standardized using comprehensive, established ontologies such as those developed by Github and EDAM. "Research Institutions" and "Funding Sources" have been standardized using the large databases of CrossRef and Wikidata, providing over 14,000 unique organization names in addition to possible organization aliases.

**Table 2: Funding Sources for Bioinformatics Tools**

| Acronym | Name | IC Abbreviation | Number of tools released since 2014 funded by IC |
|---|---|---|---|
| OD | [NIH Office of the Director (OD)](#) | OD | 17 |
| NCI | [National Cancer Institute (NCI)](#) | CA | 305 |
| NEI | [National Eye Institute (NEI)](#) | EY | 14 |
| NHLBI | [National Heart, Lung, and Blood Institute (NHLBI)](#) | HL | 117 |
| NHGRI | [National Human Genome Research Institute (NHGRI)](#) | HG | 360 |
| NIA | [National Institute on Aging (NIA)](#) | AG | 39 |
| NIAAA | [National Institute on Alcohol Abuse and Alcoholism (NIAAA)](#) | AA | 49 |
| NIAID | [National Institute of Allergy and Infectious Diseases (NIAID)](#) | AI | 72 |
| NIAMS | [National Institute of Arthritis and Musculoskeletal and Skin Diseases (NIAMS)](#) | AR | 18 |
| NIBIB | [National Institute of Biomedical Imaging and Bioengineering (NIBIB)](#) | EB | 38 |
| NICHD | [Eunice Kennedy Shriver National Institute of Child Health and Human Development (NICHD)](#) | HD | 18 |
| NIDCD | [National Institute on Deafness and Other Communication Disorders (NIDCD)](#) | DC | 3 |
| NIDCR | [National Institute of Dental and Craniofacial Research (NIDCR)](#) | DE | 22 |
| NIDDK | [National Institute of Diabetes and Digestive and Kidney Diseases (NIDDK)](#) | DK | 50 |
| NIDA | [National Institute on Drug Abuse (NIDA)](#) | DA | 45 |
| NIEHS | [National Institute of Environmental Health Sciences (NIEHS)](#) | ES | 56 |
| NIGMS | [National Institute of General Medical Sciences (NIGMS)](#) | GM | 457 |
| NIMH | [National Institute of Mental Health (NIMH)](#) | MH | 88 |
| MINHD | [National Institute on Minority Health and Health Disparities (NIMHD)](#) | MD | 8 |
| NINDS | [National Institute of Neurological Disorders and Stroke (NINDS)](#) | NS | 38 |
| NINR | [National Institute of Nursing Research (NINR)](#) | NR | 1 |
| NLM | [National Library of Medicine (NLM)](#) | LM | 103 |
| CC | [NIH Clinical Center (CC)](#) | CL | 0 |
| CIT | [Center for Information Technology (CIT)](#) | CT | 1 |
| CSR | [Center for Scientific Review (CSR)](#) | RG | 2 |
| FIC | [Fogarty International Center (FIC)](#) | TW | 2 |
| NCATS | [National Center for Advancing Translational Sciences (NCATS)](#) | TR | 17 |
| NCCIH | [National Center for Complementary and Integrative Health (NCCIH)](#) | AT | 2 |
| Total Tools Funded by NIH | | | 1434 |

**Table 2: Funding Sources for Bioinformatics Tools**

Using GROBID, funding agencies and award numbers are automatically extracted from PDF publications. Funding agency names in a publication's acknowledgements are matched to funding agency names obtained from CrossRef. Listed here are the number of tools in Aztec funded in part by each of the NIH centers, institutes, and agencies.

**Figures**

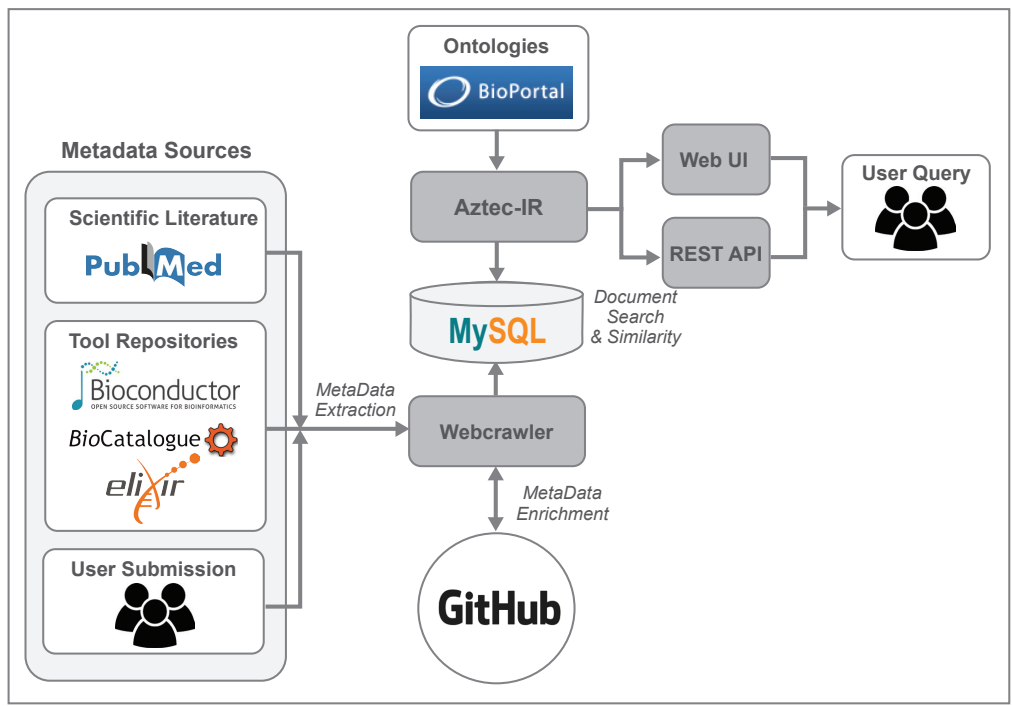

**Figure 1: Aztec Platform Architecture**

The Aztec database holds metadata on tools from four sources: (1) Tools mined from the bioinformatics literature in PubMed, (2) Tools aggregated from a network of software indexes and source code repositories, (3) Tools submitted by users or tool creators, and (4) Tools funded by the NIH BD2K initiative, which are manually entered and curated by Aztec staff.

Users interact with Aztec either through a built-in graphical interface written in HTML, CSS, and JavaScript, or directly through an application programming interface (API). Site usage data is stored in a MongoDB NoSQL database while software tool metadata is stored in a MySQL database. Document search and matching is performed using Aztec-IR, a custom built information retrieval system.

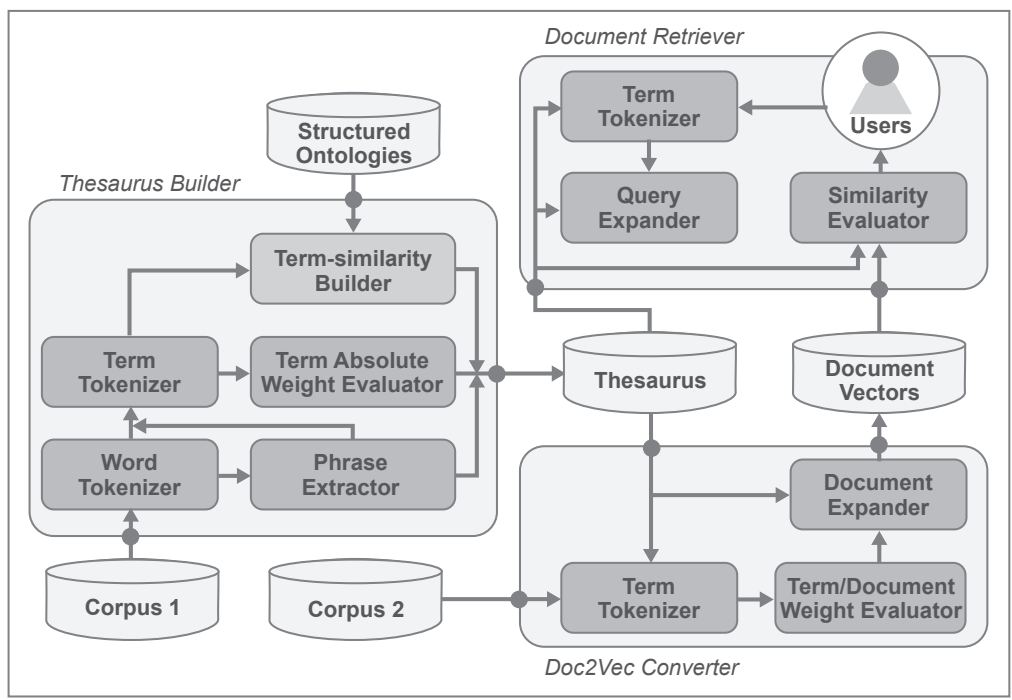

**Figure 2: Aztec-IR Workflow**

The Aztec-IR workflow can be applied to search and document similarity in domains outside of bioinformatics. To customize Aztec's search for the bioinformatics domain, Corpus 1 is built from the abstracts of bioinformatics literature on PubMed. A thesaurus is built by extracting high-occurrence phrases and identifying synonyms with high semantic similarity. Structured ontologies are used to expand and validate synonyms. Corpus 2 is built from the tool descriptions and metadata in Aztec. User queries and tool metadata are both vectorized using the thesaurus and compared for similarity. Query results are context-aware and domain relevant.

**Supplemental Figure**

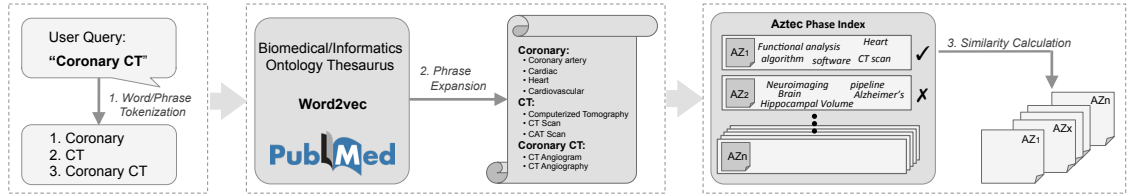

**Figure S1. Related to Figure 2. Aztec Information Retrieval (IR) Workflow**

The Aztec IR process is composed of 3 main steps. 1) A user's query is processed and tokenized into phrases. 2) The tokenized query is expanded to synonyms or similar phrases based on our Biomedical/Informatics Ontology Thesaurus. This thesaurus was created by extracting phrases trained on biomedical literature in PubMed and synonyms were generated for each extracted phrase using semantic similarity analysis. 3) Relevant results are retrieved via a similarity calculation between the phrase-expanded query and the documents in the Aztec Phrase Index. Each document in the Aztec Phrase Index represents a tool; words in the tool's metadata are tokenized and expanded similar to the methods applied to the user's query.